# SHELL MODEL PREDICTIONS FOR VERY NEUTRON-RICH NUCLEI IN $^{132}$Sn REGION : INTERESTING FEATURES OF EFFECTIVE-CHARGE REVEALED

SUKHENDUSEKHAR SARKAR and M.SAHA SARKAR*

*Department of Physics, The University of Burdwan, Golapbag, Burdwan 713104, India,
sukhendusekhar_sarkar@hotmail.com*
*Saha Institute of Nuclear Physics,1/AF Bidhan Nagar, Kolkata -- 700064, India,
mss@anp.saha.ernet.in*

ABSTRACT

Shell model studies have been done for very neutron – rich nuclei in the range $50 \leq Z \leq 55$ and $82 \leq N \leq 87$. Good agreement of the theoretical level spectra with the experimental one for N=82, 83 I and Te nuclei is shown. Then the results for three very neutron-rich nuclei $^{137}$Sn and $^{136-137}$Sb have been presented. The present calculation favour a $2^-$ ground state for $^{136}$Sb instead of $1^-$ identified through β-decay. Interesting observation about the E2 effective charges for this region has been discussed.

## 1. Introduction

Few-valence-particle neutron- rich nuclei above the strongest shell closure at $^{132}$Sn are of recent interest both experimentally as well as theoretically. Study of these close-to-drip-line nuclei, particularly the isotopes of Sn and Sb, in this mass region, are important not only for the nuclear structure, such as for the knowledge of emperical N-N interaction in the neutron-rich environment, but also for the applications in the astrophysical r-process calculations.

The model space considered for shell model calculations in this region generally assumes $^{132}$Sn as the inert core and [(proton) $\pi(1g_{7/2}, 2d_{5/2}, 2d_{3/2}, 3s_{1/2}, 1h_{11/2}$ ) and (neutron) $\nu(1h_{9/2}, 2f_{7/2}, 2f_{5/2}, 3p_{3/2}, 3p_{1/2}, 1i_{13/2})$] valence orbitals. The Hamiltonians consisting of single particle energies (spe) of the valence orbitals and the two-body matrix elements (tbme) of the residual interaction appropriate for this model space were found to be KH5082 and CW5082 [1]. Recently, realistic effective Hamiltonian based on Bonn potential and others have also been used [2, 3]. Recently we have made some modification of the CW5082 Hamiltonian in the light of recently available information on the binding energy of $^{132}$Sn and binding energies, low-lying spectra of A=134 Sn, Sb and Te isotopes and obtained two slightly different Hamiltonians named SMN and SMPN [4]. Shell model calculations using Oxbash [5] with the new Hamiltonians works remarkably well in predicting binding energies, low-lying spectra and electromagnetic transition probabilities for N=82,83 and even for N≥84 isotones of Sn, Sb, Te, I, Xe and Cs nuclei. This improvement in the predictive power [4] with the new Hamiltonians encourges us to calculate the binding energies, low-lying spectra and life-time of levels in $^{135-137}$Sn, $^{135-137}$Sb and other Te and I isotopes hitherto unknown experimentally. The low-lying spectra of $^{135}$Sn predicted by the new Hamiltonian agree [4] well with that calculated with the Hamiltonian derived from CD Bonn potential [3] as well as with a few systematic estimates [6].

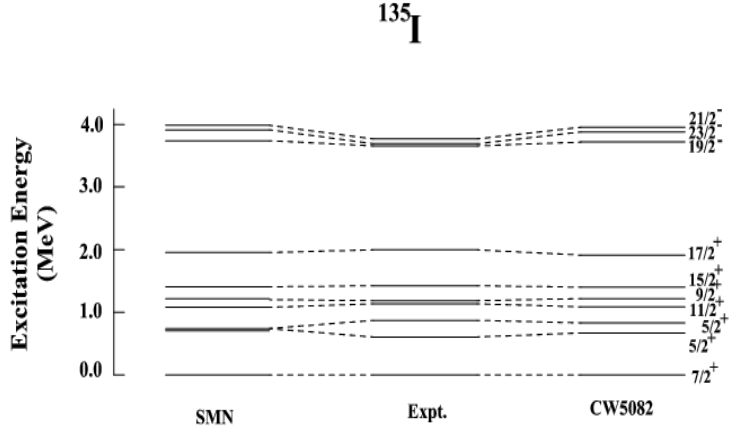

Fig. 1. *Experimental and theoretical level spectra for $^{135}$I. The results with CW5082 is taken from Ref. [4]*

## 2. Results and Discussions

The level spectra for $^{135}$I and $^{135}$Te compare well with experimental ones [7] (Figs 1 and 2). The states above 4 MeV may have components from core excitations and can not be described in this model space. The calculated energy levels and the corresponding dominant components of the wave functions are predicted for $^{137}$Sn, $^{137}$Sb in Table 1.

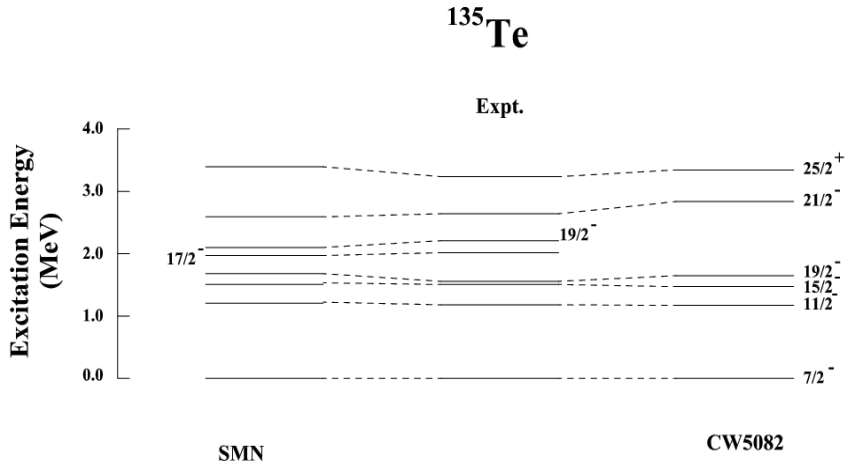

Fig. 2. *Experimental and theoretical level spectra for $^{135}$Te. The results with CW5082 is taken from Ref. [4]*

The ground state of $^{136}$Sb was argued to be 1$^-$ by Hoff et al. [8], based on its strong beta-transitions to the 0$^+$ ground and the first excited 2$^+$ states in $^{136}$Te, as well as from the analogy with the $^{212}$Bi of the

[208]Pb region, the two regions being similar [9]. The ground state of [212]Bi is 1[-]. The ground state in [136]Sb arising predominantly from the $\pi 1g_{7/2}\ \nu 2f_{7/2}^3$ configuration must have negative parity and the argument in favour of exclusion of a 0[-] possibility for the ground state is strong and plausible [8]. However, Hoff et al ruled out the possibility of a 2[-] ground state only on the ground that it demanded an exceptionally fast {$log\ f^{1u}t$ = 8.2) first forbidden unique transition to the 0[+] ground state of [136]Te. Hoff et al used, in favour of their assignment, the smallness of the unique matrix elements from the β-decay of the [134]Sb and [135]Sb. This identification has also been supported by Mineva et al [10] by their shell model calculations with KH5082 type Hamiltonians.

*TABLE I. Calculated excitation energies of yrast negative parity states in [137]Sn, and positive parity states in [137]Sb. Dominant components in the wavefunctions are also mentioned below.*

| [137]Sn | | | | [137]Sb | | |
|---|---|---|---|---|---|---|
| *I* | *Energy (MeV)* | *Wavefunction* | | *I* | *Energy (MeV)* | *Wavefunction* |
| 7/2 | 0.0(-16.046) | 68.1% $\nu 2f_{7/2}^5$ | | 7/2 | 0.0 (-24.477) | 50.5% $\pi 1g_{7/2}\nu 2f_{7/2}^4$ |
| 5/2 | 0.155 | 74.5% $\nu 2f_{7/2}^5$ | | 3/2 | 0.337 | 48.7% $\pi 1g_{7/2}\nu 2f_{7/2}^4$ |
| 3/2 | 0.380 | 79.1% $\nu 2f_{7/2}^5$ | | 1/2 | 0.511 | 47.1% $\pi 1g_{7/2}\nu 2f_{7/2}^4$ |
| 11/2 | 0.510 | 79.4% $\nu 2f_{7/2}^5$ | | 11/2 | 0.550 | 59.0% $\pi 1g_{7/2}\nu 2f_{7/2}^4$ |
| 9/2 | 0.535 | 80.7% $\nu 2f_{7/2}^5$ | | 5/2 | 0.560 | 26.2% $\pi 1d_{5/2}\nu 2f_{7/2}^4$ |
| 15/2 | 0.903 | 86.7% $\nu 2f_{7/2}^5$ | | 9/2 | 0.785 | 72.3% $\pi 1g_{7/2}\nu 2f_{7/2}^4$ |
| 1/2 | 1.338 | 63.4% $\nu 2f_{7/2}^4 3p_{3/2}$ | | 15/2 | 0.883 | 69.6% $\pi 1g_{7/2}\nu 2f_{7/2}^4$ |
| 13/2 | 1.967 | 74.2% $\nu 2f_{7/2}^4 3p_{1/2}$ | | 13/2 | 1.083 | 75.3% $\pi 1g_{7/2}\nu 2f_{7/2}^4$ |
| 17/2 | 2.305 | 73.3% $\nu 2f_{7/2}^4 1h_{9/2}$ | | 17/2 | 1.308 | 76.7% $\pi 1g_{7/2}\nu 2f_{7/2}^4$ |
| 21/2 | 2.669 | 80.3% $\nu 2f_{7/2}^4 1h_{9/2}$ | | 19/2 | 1.407 | 70.1% $\pi 1g_{7/2}\nu 2f_{7/2}^4$ |
| 19/2 | 2.682 | 79.3% $\nu 2f_{7/2}^4 1h_{9/2}$ | | 21/2 | 1.926 | 79.9% $\pi 1g_{7/2}\nu 2f_{7/2}^3 1h_{9/2}$ |
| 23/2 | 3.259 | 91.9% $\nu 2f_{7/2}^4 1h_{9/2}$ | | 23/2 | 2.051 | 77.8% $\pi 1g_{7/2}\nu 2f_{7/2}^4$ |
| | | | | 25/2 | 2.742 | 80.7% $\pi 1g_{7/2}\nu 2f_{7/2}^3 1h_{9/2}$ |

In our shell model calculation, we get a 2[-] ground state for [136]Sb with both SMN and CW5082 Hamiltonians (Fig.3). The 1[-] first excited state comes at 47 keV with the SMN Hamiltonian [4] and at 125 keV with CW5082. With KH5082 Hamiltonian, the ground state is 1[-] and 2[-] first excited state comes at an excitation energy of 126 keV. In our shell model calculations with SMN Hamiltonian, the first excited 5/2[+] state in [135]Sb is not reproduced. The experimental level energy is 282 keV, whereas the calculated value is 690 keV. Following a discussion in [2] we have reproduced this level energy in our calculation by lowering the $2d_{5/2}$ single particle energy by about 570 keV. This further improves the prediction for the other level energies of the [135]Sb spectra. With this lowered $2d_{5/2}$ spe value, calculation again predicts a 2[-] ground state and the first excited 1[-] state at 37 keV. Thus on the basis of shell model calculation, it is difficult to exclude a 2[-] possibility of the spin-parity for the [136]Sb ground state, suspecting the 1[-] identification.

The origin of the isomeric transition in $^{136}$Sb observed recently by Mineva et al [10] through 173 keV gamma is quite interesting. Mineva et al examined several probable scenarios for this isomeric transition mainly on the basis of empirical SM calculations. According to them the observed 173 keV gamma ray originate as a result of E2 transition from $4^- \to 2^-$ following the isomeric $6^- \to 4^-$ transition. Thus $6^-$ level is the probable isomer. The $6^- \to 4^-$ and $2^- \to 1^-$ gammas were not observed due to conversion and/or absorption in the aluminium catcher. Our calculation of the half-life for the excited states also favour the $6^-$ level as the isomer. From the measured half-life of 565±50 ns of the isomeric level we deduce the B(E2) values to be 187.7, 170.4 and 156.6 $e^2$ fm$^4$, respectively, with the calculated transition energy of 47 keV between $6^- \to 4^-$ levels. Effective charges deduced from this is $e_p = 1.00$ and $e_n = 1.00^{+0.06}_{-0.05}$, which is consistent for this region [4]. The observed 173 keV gamma is then originating from the prompt $4^- \to 2^-$ (gs) transition, whose calculated energy is 148 keV, only 25 keV less. A very short half-life of the $4^-$ level is obtained with the deduced set of effective charges. The neutron effective charge deduced from this isomeric transition in $^{136}$Sb appears to be slightly larger compared to that deduced from the N=84 isotones of Sn, Sb, Te nuclei [4]. It is to be noted that in deriving the set of effective charges in $^{136}$Sb we have used calculated gamma energy which may be slightly uncertain.

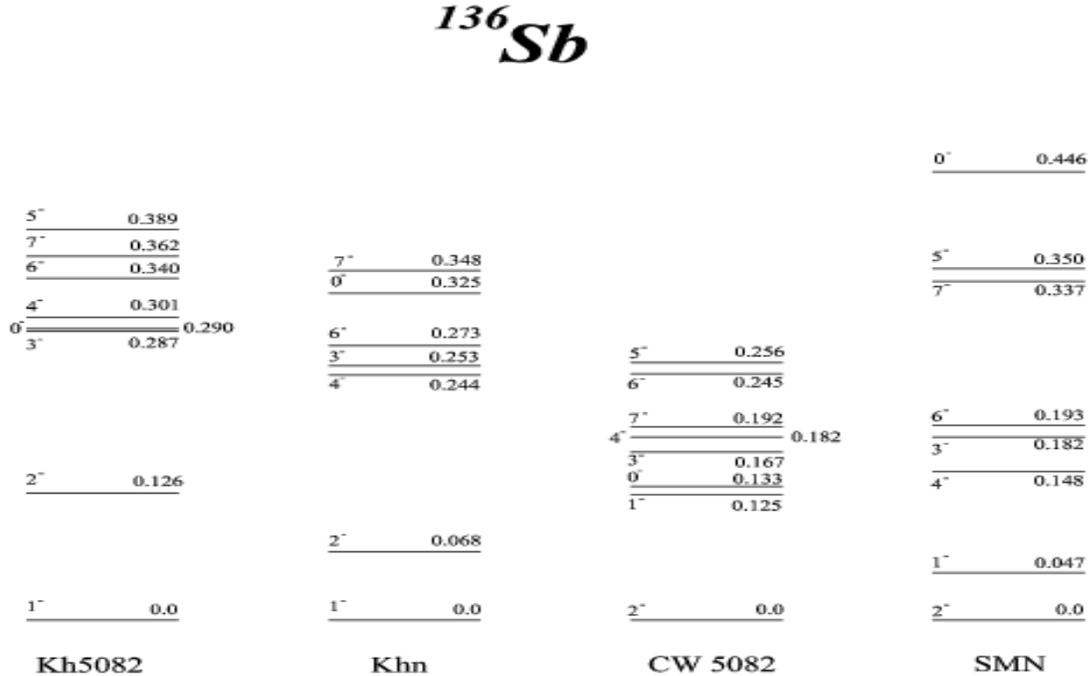

Fig. 3. *Theoretical level spectra for $^{136}$Sb with different interactions. The Khn Hamiltonian is defined in Ref. [10].*

In this context it is interesting to mention a point regarding the effective charges in this very neutron-rich close-to-dripline region. The set of E2 effective charges,(for example, $e_p = 1.47$-$1.55$, $e_n = 0.70$-$1.00$) derived from the E2 transitions in the N=82-83 nuclei can not reproduce B(E2) values in the N = 84 nuclei. For N = 84, $^{135}$Sb and $^{136}$Te nuclei a drastic reduction of proton effective charge is needed [4] to bring all the calculated B(E2) values close to the experimental ones. This is found in all the untruncated SM calculations [2, 4] over the valence space used, and is perhaps not sensitive to the parametrisation of the Hamiltonians used in this region.